# IMJENSE: Scan-specific Implicit Representation for Joint Coil Sensitivity and Image Estimation in Parallel MRI

Ruimin Feng, Qing Wu, Jie Feng, Huajun She, Chunlei Liu, Yuyao Zhang, and Hongjiang Wei

*Abstract*—**Parallel imaging is a commonly used technique to accelerate magnetic resonance imaging (MRI) data acquisition. Mathematically, parallel MRI reconstruction can be formulated as an inverse problem relating the sparsely sampled k-space measurements to the desired MRI image. Despite the success of many existing reconstruction algorithms, it remains a challenge to reliably reconstruct a high-quality image from highly reduced k-space measurements. Recently, implicit neural representation has emerged as a powerful paradigm to exploit the internal information and the physics of partially acquired data to generate the desired object. In this study, we introduced IMJENSE, a scan-specific implicit neural representation-based method for improving parallel MRI reconstruction. Specifically, the underlying MRI image and coil sensitivities were modeled as continuous functions of spatial coordinates, parameterized by neural networks and polynomials, respectively. The weights in the networks and coefficients in the polynomials were simultaneously learned directly from sparsely acquired k-space measurements, without fully sampled ground truth data for training. Benefiting from the powerful continuous representation and joint estimation of the MRI image and coil sensitivities, IMJENSE outperforms conventional image or k-space domain reconstruction algorithms. With extremely limited calibration data, IMJENSE is more stable than supervised calibrationless and calibration-based deep-learning methods. Results show that IMJENSE robustly reconstructs the images acquired at 5× and 6× accelerations with only 4 or 8 calibration lines in 2D Cartesian acquisitions, corresponding to 22.0% and 19.5% undersampling rates. The high-quality results and scanning specificity make the proposed method hold the potential for further accelerating the data acquisition of parallel MRI.**

## I. INTRODUCTION

Magnetic Resonance Imaging (MRI) is a widely used imaging technique in clinical diagnosis and research due to its safety, radiation-free, and excellent soft tissue contrast. However, MRI suffers from the main drawback of long acquisition time. Various strategies have been proposed to accelerate the acquisition of MRI images by reconstructing artifact-free MRI images from partially sampled k-space data beyond the Nyquist sampling theory. Currently, parallel MRI is used in nearly all clinical systems for scan acceleration [1, 2]. The parallel MRI reconstruction methods exploit the information redundancy of multiple receiver coils in the image domain or the k-space domain. The former requires the explicit pre-calculation of coil sensitivity maps [2-5], while the latter reconstructs the missing k-space data based on learned kernels from a fully sampled auto-calibration signal (ACS) region [1, 6-8], the structured low-rankness of multi-coil k-space data [9-11], or a combination of both [12, 13]. Compressed sensing theory offers an alternative approach to accelerate MRI. The k-space data are non-uniformly undersampled and the resulting incoherent artifacts in images can be mitigated by imposing proper constraints, such as sparsity in the transform domain [14-17] and low-rankness [18-20]. These compressed sensing-based methods have shown superior performance in 3D imaging and dynamic imaging [16, 17, 21, 22]. For the 2D Cartesian acquisition, several pseudo-random sampling patterns have been proposed to meet the requirement of incoherent undersampling [15].

In recent years, deep learning-based methods have shown promising results for MRI reconstruction. Typically, neural networks are trained in a supervised fashion via a large number of training data [23-28]. However, fully sampled ground truth data are unavailable in most practical situations, and generalization problems may arise when the inference dataset has a different distribution from the training dataset [29]. Therefore, self-supervised learning approaches have been developed to avoid the use of external fully sampled data as training labels [30]. Additionally, another category of scan-specific methods has been proposed, which employs convolution neural networks (CNN) to perform k-space interpolation or refinement for parallel MRI [31-34]. Although these methods outperform the GRAPPA at high acceleration

This paragraph of the first footnote will contain the date on which you submitted your paper for review. It will also contain support information, including sponsor and financial support acknowledgment. For example, "This work was supported in part by the U.S. Department of Commerce under Grant BS123456".

The next few paragraphs should contain the authors' current affiliations, including current address and e-mail. For example, F. A. Author is with the National Institute of Standards and Technology, Boulder, CO 80305 USA (e-mail: author@ boulder.nist.gov).

S. B. Author, Jr., was with Rice University, Houston, TX 77005 USA. He is now with the Department of Physics, Colorado State University, Fort Collins, CO 80523 USA (e-mail: author@lamar.colostate.edu).

T. C. Author is with the Electrical Engineering Department, University of Colorado, Boulder, CO 80309 USA, on leave from the National Research Institute for Metals, Tsukuba, Japan (e-mail: author@nrim.go.jp).

factors, they still depend on sufficient ACS lines to reconstruct the high-quality MR images by effectively training the CNN.

The coil sensitivities involved in most of the above methods are pre-calculated and then directly applied to the parallel MRI reconstruction model. However, accurately estimating these sensitivities is challenging from a small set of ACS regions. Errors in sensitivity maps will be amplified and lead to visible artifacts in the reconstructed image. Thus, the requirement of a larger calibration region restricts the further acceleration of k-space data acquisition. To overcome this problem, several methods that alternately update the MRI image and coil sensitivities have been proposed and have shown improved robustness to the size of the ACS region [35-44].

Recently, implicitly defined signal representations have emerged as a powerful paradigm in various fields [45-51]. This approach aims to produce a continuous internal representation of the unknown object by modeling the desired object itself as a continuous function of spatial coordinates. This function is usually parameterized by a fully connected neural network, i.e., multilayer perceptron (MLP), thus referred to as implicit neural representation (INR). The INR is learned on discrete samples by minimizing a pre-designed loss function via the gradient back-propagation algorithm. Once trained, the network implicitly encodes the continuous function, enabling the desired object to be obtained by simply querying the network with the corresponding coordinates. Implicit representation offers the following benefits: (1) the continuous nature of the representation function, which allows it to learn the internal redundancies of an object, facilitating the discovery of better solutions for physics-based problems; (2) the ease of integration with other explicit prior knowledge by designing an appropriate framework.

Based on the insights provided by implicit representation, we proposed IMJENSE, a novel framework that jointly reconstructs MRI images and coil sensitivity maps in parallel MRI. Specifically, IMJENSE models MRI images and coil sensitivities as continuous functions of spatial coordinates. The image function was parameterized by MLPs and the sensitivity map function was parameterized by polynomials. The weights in MLPs and coefficients in polynomials were simultaneously learned from partially-measured k-space data, without training databases. We evaluated IMJENSE on different datasets with varying acceleration factors and ACS sizes. The results show that IMJENSE outperforms conventional image or k-space domain reconstruction methods both qualitatively and quantitatively. With extremely limited calibration data, IMJENSE is more stable than supervised calibrationless and calibration-based deep-learning methods, allowing for higher effective acceleration rates. The main contributions of our study are:

1) IMJENSE is the first method that explores the application of INR for improving parallel MRI reconstruction.

2) IMJENSE is a scan-specific and training database-free deep learning approach. Therefore, it can be applied to different imaging modalities, acceleration factors, MRI manufacturers, and organs without necessitating a training dataset.

3) IMJENSE jointly estimates coil sensitivities and MRI images, preventing error propagation from the sensitivity maps to the reconstructed image, thus being more robust to ACS sizes.

4) IMJENSE can be easily combined with other explicit image regularizers that are commonly used in the image processing field.

## II. Related works

### A. Parallel Imaging

Sensitivity encoding (SENSE) [2] and generalized autocalibrating partially parallel acquisition (GRAPPA) [1] are the two most commonly used algorithms for parallel MRI reconstruction. SENSE formulated the reconstruction as an inverse problem in the image domain and explicitly required pre-calculated coil sensitivity maps. GRAPPA performed reconstruction in the k-space domain, where the missing k-space data were interpolated linearly using shift-invariant kernels determined from the ACS region. Additionally, some studies used CNN to interpolate or refine k-space in a scan-specific manner. For instance, Akçakaya *et al.* [31] trained nonlinear CNNs on the ACS data to replace the linear kernel in GRAPPA. This method was based on the hypothesis that the presence of noise would cause nonlinear effects on the estimation of missing k-space lines from the acquired ACS lines. Zhang *et al.* [33] utilized a residual CNN with hybrid linear and nonlinear designs to further improve the image quality and offer an interpretation for the use of nonlinear CNN. Arefeen *et al.* [32] proposed a scan-specific k-space refinement scheme that could be applied to the k-space obtained by any parallel MRI reconstruction methods to correct k-space errors and further improve reconstruction performance.

### B. Joint Estimation of Sensitivity Maps and MRI Image

In traditional algorithms, Ying *et al.* [42] proposed to alternately update the desired MRI image and coil sensitivities. The sensitivity map was parameterized by a polynomial function to reduce the unknowns to be determined. For deep learning-based methods, Jun *et al.* [41] incorporated coil sensitivity optimization into a model-based deep learning framework. Similarly, Arvinte *et al.* [35] unrolled the alternating optimization scheme into interleaved CNN blocks and data consistency blocks. Additionally, Sun *et al.* [40] employed an unrolling-based joint cross-attention network to simultaneously optimize the MRI image and sensitivity maps, guided by the already acquired fully sampled intra-subject scan of different contrast. Wang *et al.* [37] developed a sensitivity estimation module to refine sensitivity maps during image reconstructions. Notably, this method utilized the entire measured k-space data rather than limiting to ACS lines, to estimate more accurate sensitivity maps. These methods trained the network in an end-to-end supervised manner, while Gan *et al.* [44] developed a self-supervised architecture that performed joint image reconstruction and sensitivity calibration without ground truth data.

### C.Implicit Neural Representation

The idea of INR first arose in view synthesis and achieved remarkable results. Mildenhall *et al.* [47] proposed to represent a 3D scene as neural radiance fields using a fully connected network (NeRF). NeRF takes a 5D coordinate including spatial location $(x, y, z)$ and viewing direction $(\vartheta, \phi)$ as input and outputs the corresponding volume density and emitted radiance. The network was optimized using a set of ground truth 2D images with known camera poses. Once optimization was completed, unobserved views could be synthesized by querying the network with the corresponding 5D coordinates.

Meanwhile, researchers found that deep networks directly operating on the input coordinates would result in poor performance in representing high-frequency variations [52]. To overcome this problem, the input coordinates were encoded into a higher dimensional space before passing them to the network. Different encoding functions were proposed to make the networks better learn high-frequency information, such as position encoding [47], Fourier feature map [53], and radial encoding [49]. In addition, Sitzmann *et al.* [54] proposed sinusoidal representation networks (SIREN) that leveraged periodic activation functions, i.e., sine function, to make MLP more capable of modeling fine details of a signal and its derivatives. Recently, parametric encodings were proposed [55, 56], which contained learnable parameters and used a sparse data structure. These encoding paradigms permit the use of smaller networks and speed up the training process.

In the field of MRI acceleration, Shen *et al.* [46] proposed implicit Neural Representation learning with Prior embedding (NeRP) to reconstruct an image from the undersampled measurements. However, NeRP requires a fully sampled prior image from a previous scan on the same subject. Therefore, the clinical application of NeRP is limited to the case where long-term longitudinal scanning is performed. Moreover, NeRP for MRI acceleration only focused on the radial k-space sampling pattern.

## III.Methodology

### A.Problem Formulation

#### 1)Inverse problem in MRI reconstruction

In the multi-channel MRI, each receiver coil partially samples the k-space. The undersampled k-space signal of the $jth$ coil, $S_j$, is given by:

$$S_j = \mathbf{E}_j I + n_j,\tag{1}$$

where $I$ represents the vectorized image to be reconstructed and $n_j$ is the noise. $\mathbf{E}_j = \mathbf{MFC}_j$ is the forward physical model, where $\mathbf{C}_j$ denotes the diagonalized sensitivity map matrix of the $jth$ coil, $\mathbf{F}$ denotes the Fourier transform matrix and $\mathbf{M}$ is the diagonalized sampling mask. Recovering $I$ from the undersampled k-space data is typically formulated as an optimization problem:

$$\arg\min_I \frac{1}{2}\sum_{j=1}^{c}\left\|S_j - \mathbf{E}_j I\right\|_2^2 + \lambda\mathcal{R}(I),\tag{2}$$

where $c$ is the total number of receiver coils, and the first term imposes data consistency with the measured signals. $\mathcal{R}(\cdot)$ is the regularization term that imposes prior knowledge on the reconstructed image $I$ and $\lambda$ balances the contributions of these two terms. In most current methods, the image $I$ is solved by given knowledge of the acquired k-space signal and the sensitivity maps that are pre-estimated using traditional algorithms, such as ESPIRiT [57]. With a smaller ACS size, any inaccuracy in the independently estimated sensitivity maps will be introduced into the subsequent image reconstruction step and degrade the quality of the reconstructed results. One possible solution is to simultaneously estimate the coil sensitivity maps and the desired MRI image. However, (1) will become extremely ill-posed in this case.

#### 2)Polynomial representation of sensitivity maps

In this study, we modeled the sensitivity map as a polynomial function $f_\varphi$ of the spatial coordinates $(x, y)$ to further reduce the unknowns in coil sensitivity maps [42]. Let $C_j(x, y)$ be the value of the $jth$ coil sensitivity at the coordinate $(x, y)$, then it can be expressed as:

$$C_j(x, y) = \sum_{p=0}^{N}\sum_{q=0}^{N}\varphi_{p,q,j}x^p y^q,\tag{3}$$

where $\varphi_{p,q,j}$ is the unknown polynomial coefficient to be estimated, and $N$ represents the order of the polynomial. The polynomial representation will naturally produce smooth values that conform to the characteristics of the sensitivity map. In addition, due to the low-dimension nature of the sensitivity map, a lower polynomial order is sufficient. Therefore, the unknowns in coil sensitivity maps are significantly reduced, making it possible for joint optimization of sensitivity maps and the MRI image. It should be mentioned that since the coil sensitivities are complex values, the real and imaginary parts are separately modeled by the polynomials in this study. Therefore, the total number of parameters for $c$ sensitivity maps is $2cN^2$.

#### 3)Implicit neural representation of an MRI image

To learn the internal continuity, we modeled the MRI image intensity as a continuous function of the spatial coordinates $(x, y)$. The MLP was used to parameterize this continuous function as adopted in the field of INR, i.e., $MLP_\theta(x, y)$, where $\theta$ denotes the weights in the MLP to be optimized. Let $I(\theta)$ and $\mathbf{C}_j(\varphi)$ be the discretized image vector and sensitivity map matrix after sampling the $MLP_\theta$ and $f_\varphi$ at the center location of the pixels, respectively. Then the joint reconstruction problem is written as:

$$\arg\min_{\theta,\varphi} \frac{1}{2}\sum_{j=1}^{c}\left\|S_j - \mathbf{MFC}_j(\varphi)I(\theta)\right\|_2^2 + \lambda\mathcal{R}[I(\theta)].\tag{4}$$

Therefore, this problem has been converted into optimizing parameters in the MLP and polynomials by searching for the solution over the function space they represent, instead of directly operating on the desired images. Since both the MLP and polynomials learn continuous functions, they implicitly incorporate the continuity prior in the represented images, resulting in suppressed noise and artifacts.

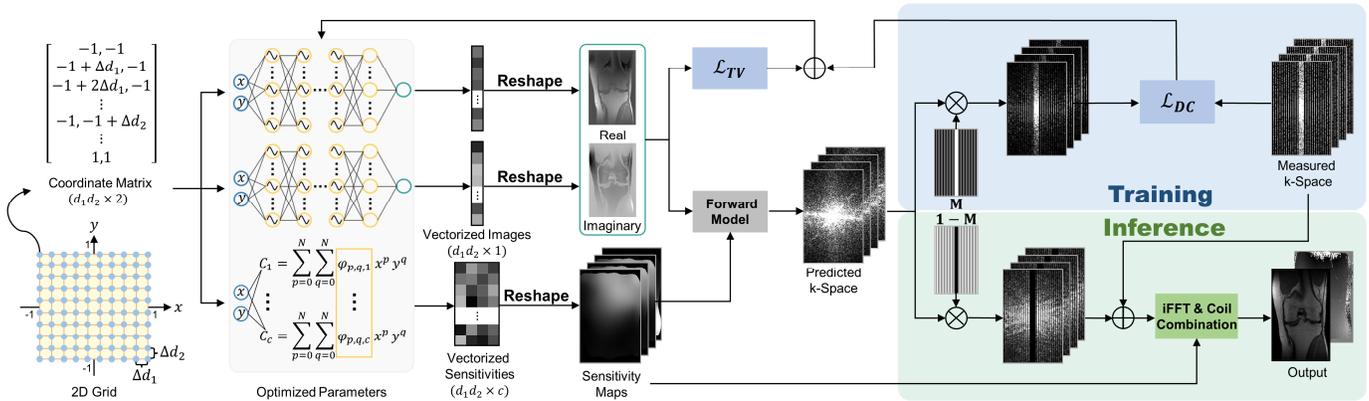

**Fig. 1.** Overview of the proposed method. Pixel coordinates are determined by their positions, with $\Delta d_1$ and $\Delta d_2$ denoting intervals between adjacent pixels along the $x$ and $y$ directions, $d_1$ and $d_2$ representing the total number of pixels along these directions. These coordinates are stacked into a matrix that serves as input for two MLPs and polynomials, producing vectorized real and imaginary components of an MRI image, along with coil sensitivities. These vectors are then reshaped to create the 2D images and sensitivity maps, which are used for k-space signal prediction. During the training, the MLP weights and the polynomial coefficients are simultaneously optimized by minimizing data consistency loss $\mathcal{L}_{DC}$ and total variation loss $\mathcal{L}_{TV}$. When inferring, an additional step is adopted to enforce the data consistency of k-space.

### B. Overall Framework of IMJENSE

Fig. 1 illustrates the overview of the proposed method. Each pixel is encoded with a unique coordinate based on its position, where $\Delta d_1$ and $\Delta d_2$ represent the intervals between adjacent pixels along the $x$ and $y$ directions, $d_1$ and $d_2$ are the total number of pixels along these directions, thus $\Delta d_1 = 2/(d_1 - 1)$ and $\Delta d_2 = 2/(d_2 - 1)$. These coordinates are stacked along the first dimension to form a matrix with the size of $d_1 d_2 \times 2$, where the first dimension represents batches, and the second dimension represents the input channel. Then, the coordinate matrix is fed into two MLPs and polynomials to produce the vectorized real and imaginary components of an MRI image, as well as the coil sensitivities. Subsequently, these vectors are reshaped to create the 2D images and sensitivity maps, which are used to predict the k-space signal via the forward physical model. During training, the weights in the MLPs and coefficients in the polynomials are optimized simultaneously by minimizing the loss function. When inferring, an additional step is adopted to enforce k-space consistency by replacing the estimated k-space data with the acquired measurements.

#### 1) Sinusoidal representation network

As mentioned in Section II, previous studies have shown that a regular MLP is biased towards low-frequency functions and poorly represents high-frequency information. To better represent the details and fine structures of an image, we adopted the sinusoidal representation network (SIREN) [54], which is essentially an MLP but with the periodic sine function as the activation function. To enhance network training efficiency, the original authors of SIREN proposed a specific initialization strategy. Briefly, the weights of all layers except the first layer of the network were initialized to follow a uniform distribution of $U(-\sqrt{6/n}, \sqrt{6/n})$, where $n$ is the number of features of the layer input. This initialization guarantees that the output distribution of each sine activation function remains consistent regardless of the number of layers. For the first layer, the weights were initialized by a uniform distribution of $U(-1/n, 1/n)$ and then multiplied by a factor $w_0$, where $w_0$ controls the spatial frequency of the first layer, which determines the network's ability to represent high-frequency information. In this study, $w_0$ is dataset-dependent and will be fine-tuned for different datasets.

#### 2) Training and inference

The weights in SIREN and polynomial coefficients were simultaneously trained by minimizing the following loss function:

$$\mathcal{L}_{tot} = \mathcal{L}_{DC} + \lambda \mathcal{L}_{TV}, \qquad (5)$$

where $\mathcal{L}_{DC}$ imposes data consistency with the partially measured k-space data, $\mathcal{L}_{TV}$ represents the total variation regularizer:

$$\mathcal{L}_{DC} = \sum_{j=1}^{c} \left\| S_j - \mathbf{MFC}_j(\varphi)I(\theta) \right\|_1, \qquad (6)$$

$$\mathcal{L}_{TV} = \|GI(\theta)\|_1, \qquad (7)$$

where $G$ is the gradient operator. In each iteration, the predefined spatial coordinates from the entire image were fed into the network to generate the complete MRI image and coil sensitivity maps. This enabled the calculation of the loss function. Subsequently, $\varphi$ and $\theta$ were updated simultaneously through a gradient step.

After completion of the training process, the neural networks learn a representation of the MRI image, which enables them to produce a reconstructed image directly. To further improve the results, an additional step that enforces the data consistency of k-space is recommended. As shown in the inference module of Fig. 1, the measured k-space data are used to replace the predicted k-space to enforce data consistency. Then, the individual coil images are obtained by applying the 2D inverse Fourier transform to these composite k-space data and are combined using the adaptive coil combination method [58].

### C. Implementation Details

SIREN is a fully connected network, where each neuron in the previous layer connects to all neurons in the subsequent layer. In this study, the SIREN network consists of one input layer, six hidden layers, and one output layer. Each hidden layer contains 256 neurons, and all layers, excluding the final one, utilize the periodic sine activation function as mentioned. The input layer receives the 2D coordinate and generates 256 features, which are then passed to the hidden layer. The output layer produces the image intensity corresponding to the given input coordinate. The spatial coordinates were normalized to the range of [-1, 1]. The order of the polynomial function was set to 15, and the polynomial coefficients were initialized with a normal distribution using a constant random seed. SIREN and polynomials were trained using two separate Adam optimizers [59]. The initial learning rate for SIREN was $1 \times 10^{-4}$, which decayed by a factor of 0.8 every 500 iterations, whereas the initial learning rate for polynomials was 0.1, which decayed by a factor of 0.5 every 500 iterations. The total number of iterations was set to 1500.

The proposed method was implemented using PyTorch 1.10.2 in Python 3.9 on a workstation with two Intel Xeon Platinum 8249C CPUs @ 2.10 GHz with 256 GB RAM and an NVIDIA GeForce RTX 3090 GPU with 24 GB memory. The code for training and testing is available at: https://github.com/AMRI-Lab/IMJENSE.

## IV. EXPERIMENTS

### A. Datasets

In this study, the following datasets were used to evaluate the performance of the proposed method.

#### 1) 15-channel knee dataset

The knee data were obtained from the NYU fastMRI Initiative database with approval from the New York University School of Medicine Institutional Review Board [60, 61]. This dataset was fully sampled with a 15-channel knee coil using a 2D turbo spin-echo sequence without fat suppression. A total of 524 slices from 15 subjects were used to evaluate the performance of different approaches. To train the supervised methods, an additional 538 slices from 15 subjects were selected as the training dataset.

#### 2) 2-channel macaque brain dataset

A macaque brain was collected with the approval of the Life Sciences Ethics Committee, Institute of Neuroscience, Chinese Academy of Sciences. This sample was scanned on a 9.4T Bruker scanner equipped with a 2-channel receiver coil. A 3D multi-echo gradient recalled echo (GRE) sequence was conducted with the following parameters: matrix size = 710×840×512, voxel size = 0.1×0.1×0.1 mm³, $TE_1/TE_2/TE_3$ = 5.7/15.2/24.7 ms, TR = 75 ms, and flip angle = 35°. The 3/4 partial Fourier was used to accelerate acquisition. The raw k-space data were 1D inverse Fourier transformed along the fully sampled slice direction, and ten slices were randomly selected for subsequent analyses (matrix size=710×840).

#### 3) 32-channel human brain dataset

Nine subjects were scanned using a 3.0T scanner (uMR790, United Image Healthcare, Shanghai, China) equipped with a 32-channel head receiver coil. The experiment was approved by the Shanghai Jiao Tong University Human Ethics Committee and the informed consent was signed before scanning. Data were acquired using a 3D Magnetization Prepared Rapid Gradient Echo (MPRAGE) sequence with the following parameters: matrix size = 240×236×160, voxel size = 1×1×1 mm³, TR/TE/TI = 2100/3.1/750 ms, flip angle = 10°, bandwidth = 250 Hz/pixel. The raw k-space data were 1D inverse Fourier transformed along the fully sampled slice direction. A total of 320 slices from 2 subjects were used to test the performance of different methods, while the remaining 7 subjects served as the training dataset for the supervised methods.

#### 4) 20-channel lesion dataset

A brain MRI dataset containing lesions was downloaded from the NYU fastMRI Initiative database. This dataset was a 20-channel fully sampled k-space scanned using the Fluid-attenuated Inversion Recovery (FLAIR) sequence with the matrix size = 320×320. Hyperintensities within the white matter of the MRI image were identified as the lesion region. An additional 484 slices from 31 subjects in the fastMRI FLAIR brain dataset were used for training the supervised methods.

To evaluate both the MRI magnitude and phase images, the reconstructed k-space data of individual channels were combined using the adaptive coil combination method [58] to generate the complex MRI image. For the macaque brain dataset, the original 3/4 k-space data were first zero-padded prior to performing the coil combination.

### B. Performance Evaluation

IMJENSE was compared with several conventional algorithms, including GRAPPA [1], JSENSE [42], Nonlinear inversion (NLINV) [38], L1-ESPIRiT [57], LORAKS [10, 62], and SHLR-SV [13], as well as two supervised deep-learning methods, H-DSLR [63] and MoDL [24]. L1-ESPIRiT is a calibration-based algorithm that requires pre-estimated sensitivity maps. For a fair comparison between the proposed implicit representation and the traditional direct representation, total variation was selected as the regularization term of L1-ESPIRiT. SHLR-SV is a state-of-the-art conventional method that combines the separable low-rank Hankel regularization and self-consistency of k-space. H-DSLR is a calibrationless deep-learning method that relies on large-scale training data, so we evaluated its performance on the knee dataset, human brain dataset, and lesion dataset. MoDL is a calibration-based deep-learning method that uses pre-calculated coil sensitivity information. Since a comparison between H-DSLR and MoDL was already conducted in the H-DSLR study, we compared MoDL under different ACS sizes to demonstrate the benefits of our proposed joint estimation scheme.

To quantitatively evaluate the reconstruction results of various methods, we computed two commonly used metrics, peak signal-to-noise ratio (PSNR) and structural similarity index (SSIM). Moreover, for methods that perform reconstruction in the k-space domain, including GRAPPA, LORAKS, SHLR-SV, H-DSLR, and IMJENSE, we presented the reconstructed magnitude k-space and its corresponding errors relative to the fully sampled k-space. This enables an intuitive comparison of the ability of each method to recover the missing k-space data.

**Table 1.** Summary of datasets, acceleration factors, ACS sizes, as well as the corresponding undersampling rates used in each experiment of Section IV.D.

| Experiments | Datasets | R and ACS | Undersampling Rates |
|---|---|---|---|
| (1) | 15-channel FastMRI knee dataset | R=4, ACS=24 | 29.9% |
| | | R=5, ACS=24 | 25.3% |
| (2) | 2-channel macaque brain dataset | R=2, ACS=100 | 56.0% |
| (3) | 32-channel human brain dataset | R=5, ACS=4, 8, 12, 16, 32 | 22.0%, 22.9%, 24.6%, 25.4%, 31.4% |
| (4) | 32-channel human brain dataset | R=5, ACS=8, 12, 16, 20, 24, 28, 32 | 22.9%, 24.6%, 25.4%, 27.1%, 28.8%, 29.7%, 31.4% |
| | | R=6, ACS=8, 12, 16, 20, 24, 28, 32 | 19.5%, 21.2%, 22.9%, 23.7%, 25.4%, 27.1%, 28.0% |
| (5) | 20-channel lesion dataset | R=4, ACS=8 | 26.9% |
| (6) | 15-channel FastMRI knee dataset | R=4, ACS=12, 24 | 27.5%, 29.9% |
| (7) | 15-channel FastMRI knee dataset | R=4, ACS=24, 32 | 29.9%, 31.5% |
| | | R=5, ACS=24, 32 | 25.3%, 27.2% |

## C. Hyperparameter Tuning of IMJENSE

IMJENSE involves two important hyperparameters, $\lambda$ and $\omega_0$. These hyperparameters were fine-tuned using only a small set of training data from each dataset and then applied to the remaining data in the dataset. To select optimal hyperparameters, we employed the Bayesian optimization method [64], which is one of the well-established techniques used for hyperparameter tuning in machine learning. Unlike grid search [65], which attempts all possible combinations of hyperparameters, Bayesian search chooses its parameter in each iteration based on the score obtained in the previous rounds. Through assessing hyperparameters with more favorable outcomes based on prior results, Bayesian search narrows down the search space and thus allows for the finding of the optimal settings in fewer iterations. Consequently, Bayesian optimization is more efficient when searching within a wide parameter space. In this study, we used the mean PSNR relative to the ground truth as the score to evaluate the performance under different hyperparameter settings. For all the datasets, the search range for $\omega_0$ ranged from 10 to 50, and the search range for $\lambda$ ranged from 0 to 100. Bayesian optimization was conducted over a total of 24 iterations, with 4 randomly selected initial points. It should be noted that Bayesian optimization selects continuous parameters, so the obtained $\omega_0$ was rounded to the nearest integer, and $\lambda$ was rounded to one decimal place.

### D. Experiment Design

To evaluate the proposed method, these datasets were retrospectively uniformly undersampled with different acceleration factors and ACS sizes. Specifically, we conducted the following experiments: (1) We tested the performance of IMJENSE and compared methods on the 15-channel knee dataset with a regular ACS size (24 ACS lines). (2) We compared the robustness of these methods to data acquired with fewer channels using the 2-channel macaque brain dataset. (3) To demonstrate the effectiveness of the proposed joint estimation scheme, we employed ESPIRiT to pre-estimate the coil sensitivity maps with varying ACS lines independently for IMJENSE. Subsequently, we compared the MRI image reconstructed by IMJENSE to that reconstructed by the same framework using the pre-estimated coil sensitivity maps calculated by ESPIRiT, termed IMJENSE-J. (4) We applied IMJENSE and compared methods to the human brain dataset with different acceleration factors and ACS lines to test the robustness of IMJENSE to ACS sizes. (5) We assessed the ability of these methods to detect abnormal image intensities using the 20-channel lesion data. (6) We compared IMJENSE and MoDL at ACS=12 and 24 to demonstrate the superiority of IMJENSE over the calibration-based supervised deep-learning method. (7) We conducted ablation experiments to illustrate the effectiveness of each configuration in the proposed method. We designed different variants by replacing the SIREN network with the traditional MLP (i.e., with ReLU activation function, termed IMJENSE-Sine), using the traditional MLP with position encoding (termed IMJENSE-Sine+PE), ablating the TV loss function (termed IMJENSE-TV), and ablating the k-space consistency step (termed IMJENSE-KC). The datasets, acceleration factors, and ACS sizes used in each experiment are summarized in Table 1.

## V. Results

### A. Results of Hyperparameter Tuning

Fig. 2 shows the hyperparameters discovered by Bayesian optimization during the parameter tuning for the FastMRI knee dataset. It can be seen that the search primarily concentrates within the intervals of [20, 40] for $\omega_0$ and [0, 20] for $\lambda$. The highest PSNR value is achieved at $\omega_0$=31 and $\lambda$=3.8. A similar procedure was employed to fine-tune the hyperparameters for other datasets. Table 2 summarizes the specific hyperparameter configurations for each dataset.

**Table 2.** The hyperparameter configurations for each dataset.

| | Knee dataset | Macaque brain dataset | Human brain dataset | Lesion dataset |
|---|---|---|---|---|
| $\omega_0$ | 31 | 24 | 17 | 22 |
| $\lambda$ | 3.8 | 36.7 | 1.5 | 4.2 |

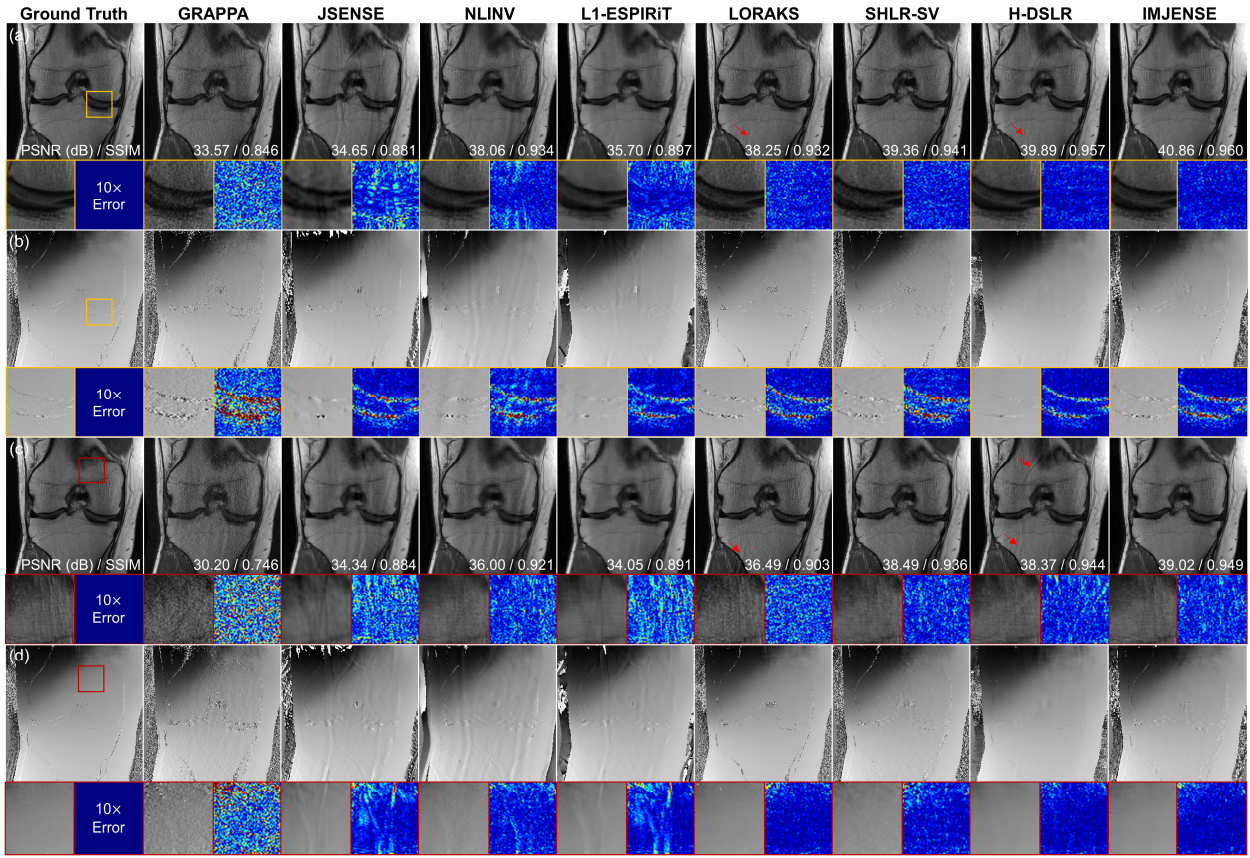

**Fig. 3.** Comparisons of different methods on the 15-channel knee dataset with 24 ACS lines at (a-b) R=4 and (c-d) R=5. The zoomed-in images show that IMJENSE effectively removes noise and artifacts in the magnitude and phase images and exhibits smaller differences relative to the ground truth. The red arrows point to the artifacts in the magnitude images reconstructed by LORAKS and H-DSLR. Quantitative evaluation metrics (PSNR and SSIM) are reported below each image. The 10× error maps are displayed for better visualization.

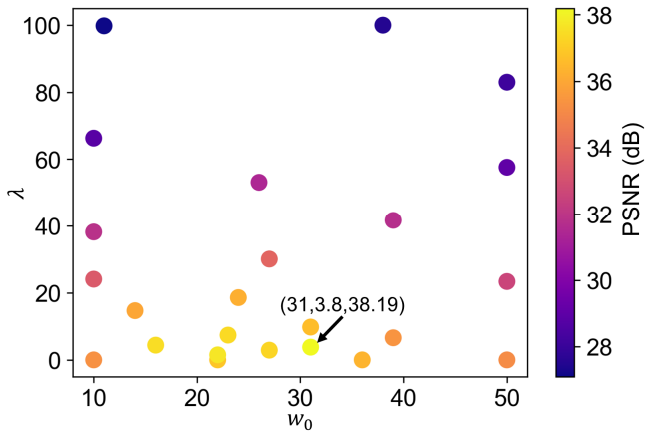

**Fig. 2.** Visualization of hyperparameter search using Bayesian optimization on the FastMRI knee training dataset.

## B. Comparisons on the Knee Dataset

Fig. 3 displays a representative knee slice reconstructed by different methods with 24 ACS lines. At R=4, SHLR-SV and IMJENSE effectively eliminate noise and aliasing artifacts in both magnitude and phase images, which are prominent in the reconstructed results of GRAPPA, JSENSE, and NLINV, as illustrated by the zoomed-in images in Figs. 3(a) and 3(b). L1-ESPIRiT gives over-smoothed results. Similar results are observed when the acceleration factor increases to 5 (Figs. 3(c) and 3(d)). Although LORAKS and H-DSLR display visually comparable performance to IMJENSE, they are unable to effectively remove artifacts in the magnitude image, as indicated by the red arrows. Notably, IMJENSE achieves a PSNR of 40.86 dB and an SSIM of 0.960 at R=4. At R=5, IMJENSE achieves a PSNR of 39.02 dB and an SSIM of 0.949. Fig. 4 displays the k-space of the data in Fig. 3 recovered by different methods, along with the error map relative to the fully sampled k-space data. SHLR-SV and IMJENSE exhibit smaller deviations than GRAPPA, LORAKS, and H-DSLR.

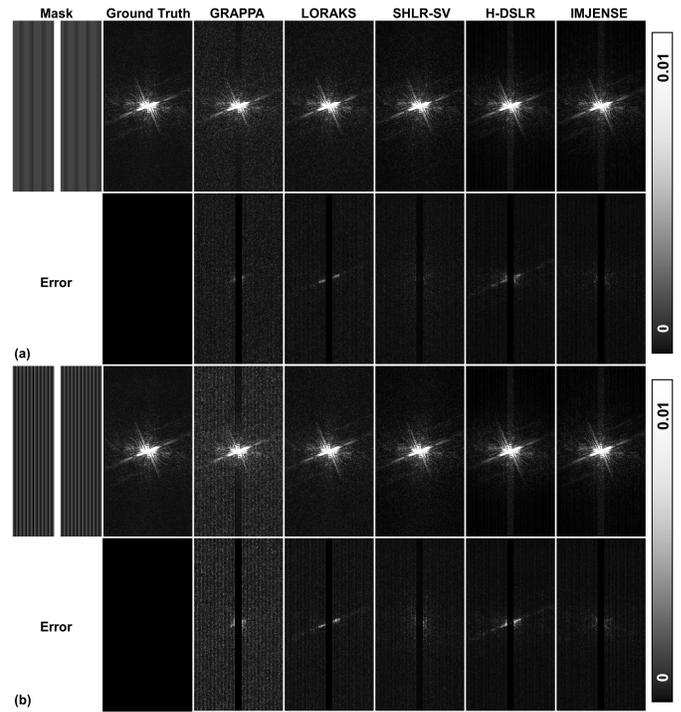

**Fig. 4.** The recovered k-space and the corresponding error map relative to the fully sampled k-space on the knee dataset with 24 ACS lines at (a) R=4 and (b) R=5. The magnitude k-space and errors from the first channel are displayed with a range of [0, 0.01] for better visualization.

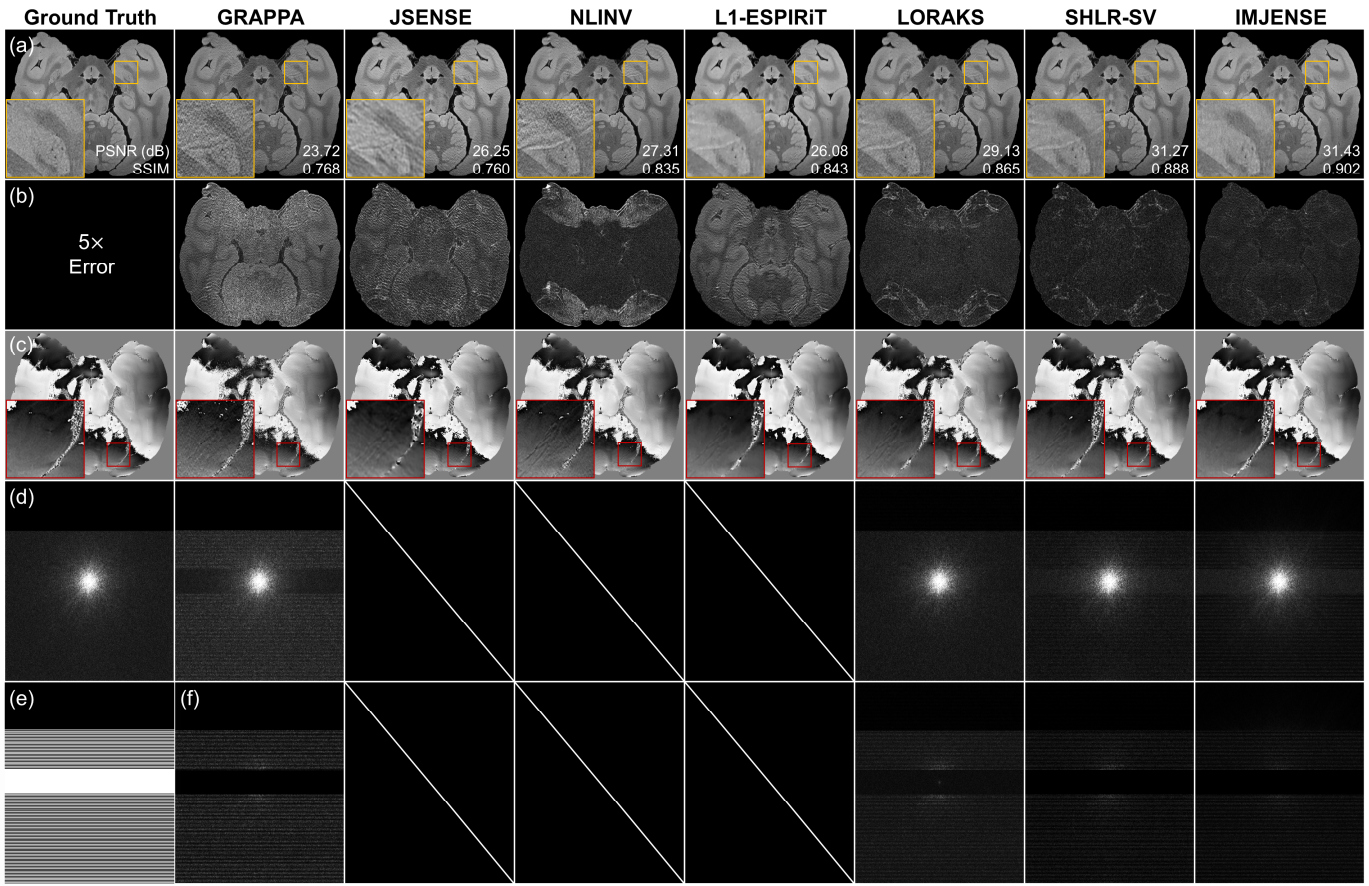

**Fig. 5.** Comparisons of different methods on the 2-channel macaque brain dataset at R=2 and ACS=100. (a) and (c) The reconstructed magnitude and phase images. Zoomed-in images show that IMJENSE removes artifacts that are obvious in the reconstructed results of the compared methods. (b) The 5× errors of the reconstructed magnitude images relative to the ground truth. (d) and (f) The magnitude of the recovered k-space and the corresponding errors relative to the fully-sampled k-space data. (e) Undersampling mask.

Table 3 reports the quantitative evaluation metrics analyzed on 524 slices from 15 subjects. IMJENSE achieves the highest PSNR of 38.27 dB and SSIM of 0.950 at R=4, as well as the highest PSNR of 36.54 dB and SSIM of 0.941 at R=5.

### C.Comparisons on the Macaque Brain Dataset

Fig. 5 presents the reconstruction results of different methods on the macaque brain dataset at R=2 and ACS=100. IMJENSE effectively removes artifacts from both magnitude and phase images that can not be suppressed by the other compared methods, as illustrated in the zoomed-in images in Figs. 5(a) and 5(c). Additionally, IMJENSE exhibits lower errors in terms of the recovered k-space relative to the fully sampled data, as shown in Fig. 5(f). Table 4 summarizes the statistical values of PSNR and SSIM calculated on 10 randomly selected slices.

IMJENSE achieves the highest PSNR of 31.71 dB and the highest SSIM of 0.899.

**Table 3.** Comparison of PSNR and SSIM obtained by different methods on 524 slices from the FastMRI knee testing dataset. The best results are in bold. Data are presented as Mean±Standard deviation.

|  | R=4 | | R=5 | |
|---|---|---|---|---|
|  | PSNR | SSIM | PSNR | SSIM |
| GRAPPA | 33.63±5.19 | 0.848±0.104 | 30.02±4.81 | 0.781±0.118 |
| JSENSE | 32.71±3.52 | 0.888±0.084 | 31.97±3.69 | 0.888±0.084 |
| NLINV | 35.71±4.74 | 0.889±0.097 | 34.07±4.54 | 0.874±0.100 |
| L1-ESPIRiT | 33.70±3.91 | 0.910±0.076 | 32.31±3.93 | 0.902±0.077 |
| LORAKS | 36.92±4.00 | 0.910±0.080 | 35.34±4.18 | 0.884±0.090 |
| SHLR-SV | 36.96±4.81 | 0.905±0.088 | 35.60±4.51 | 0.893±0.092 |
| H-DSLR | 37.03±4.20 | 0.949±0.057 | 36.02±3.98 | 0.936±0.062 |
| IMJENSE | **38.27±4.18** | **0.950±0.068** | **36.54±4.12** | **0.941±0.070** |

### D.Effects of Joint Estimation

Fig. 6 compares the ability of ESPIRiT and IMJENSE to estimate sensitivity maps. ESPIRiT generates inaccurate coil sensitivity maps with only 4 ACS lines compared to that with

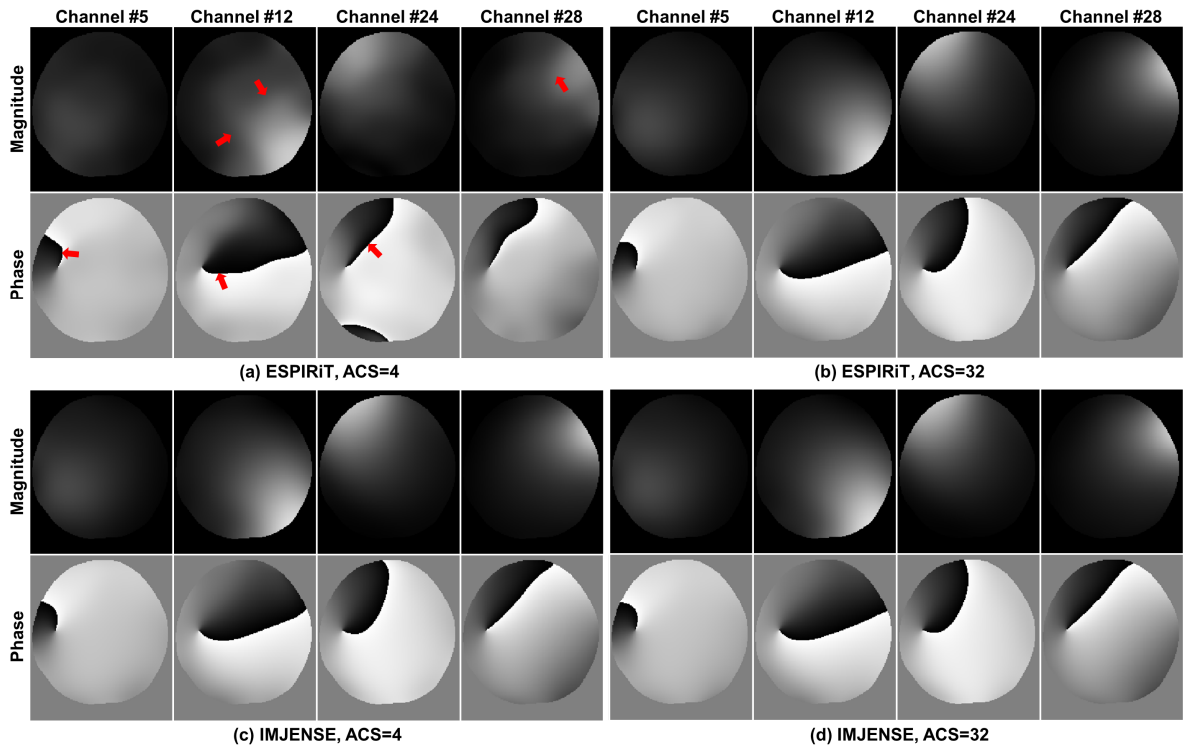

**Fig. 6.** Magnitude and phase coil sensitivity maps from the human brain dataset estimated by (a-b) the ESPIRiT algorithm and (c-d) IMJENSE with ACS=4 and ACS=32, respectively. The coil sensitivity maps from four representative channels are presented.

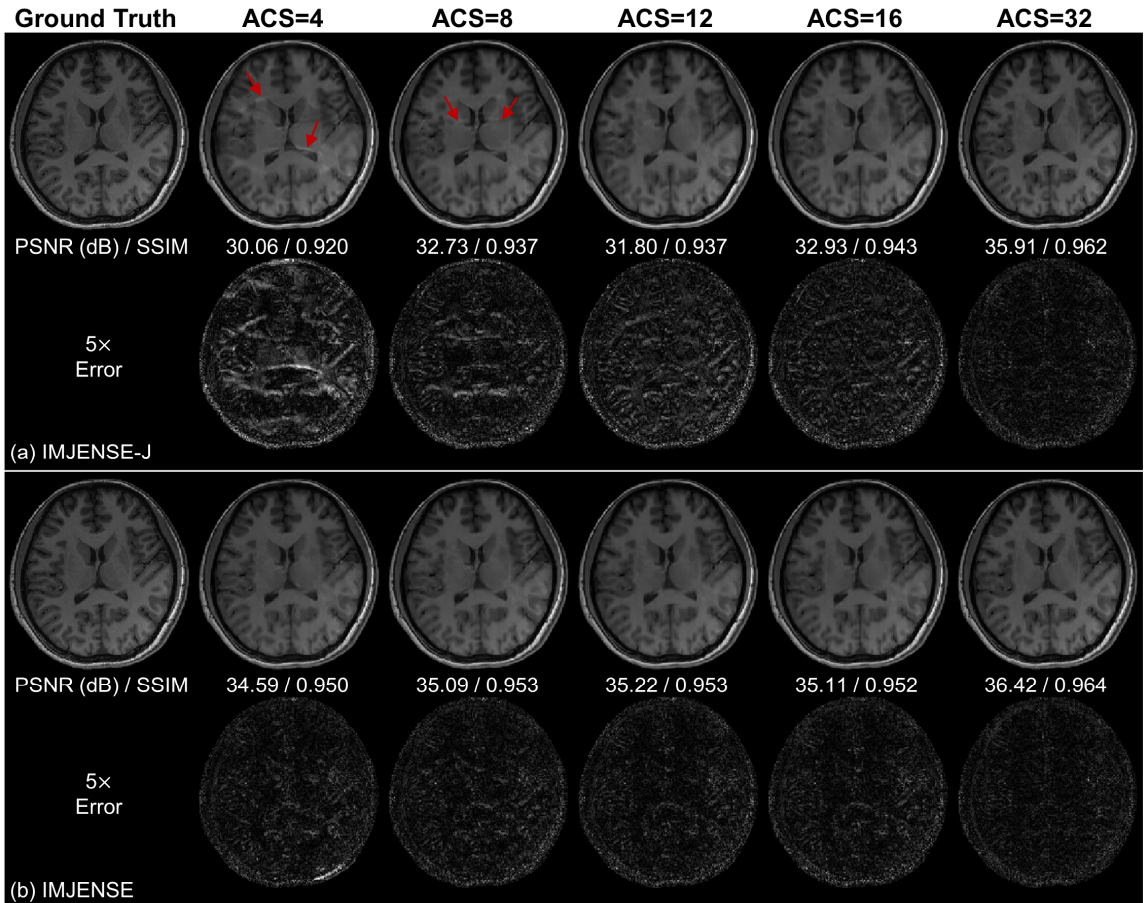

**Fig. 7.** The MRI images reconstructed by (a) IMJENSE-J (i.e., IMJENSE with the pre-estimated coil sensitivity maps by ESPIRiT) and (b) IMJENSE on the human brain dataset at R=5. Red arrows point to the artifacts in the IMJENSE-J results with fewer ACS lines. The PSNR (dB) and SSIM are reported below each image.

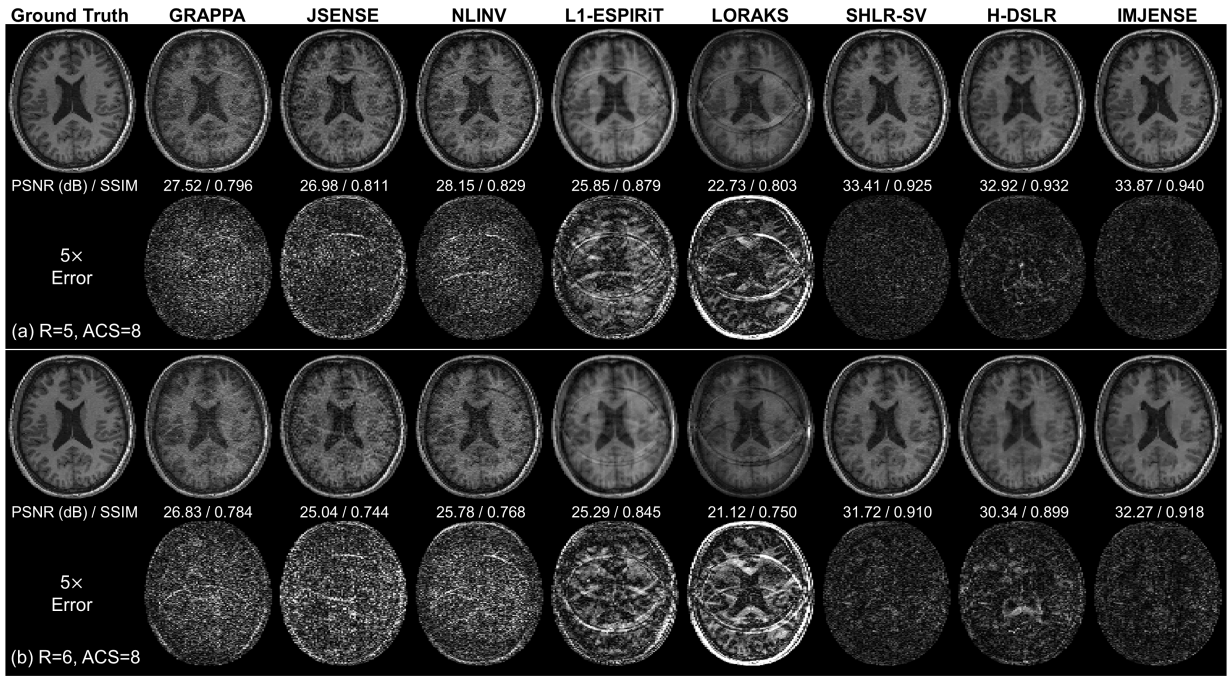

**Fig. 8.** Comparisons of different methods on the 32-channel human brain dataset with 8 ACS lines at (a) R=5 and (b) R=6. Quantitative evaluation metrics are reported below each image. IMJENSE achieves the highest PSNR and SSIM.

32 ACS lines as indicated by the red arrows, due to the insufficient number of calibration data. In contrast, IMJENSE provides consistently superior coil sensitivity estimations with both 4 and 32 ACS lines, as shown in Figs. 6(c) and 6(d). Fig. 7 compares the MRI image reconstructed by IMJENSE-J and IMJENSE at R=5. Visually, the results of IMJENSE-J exhibit obvious aliasing artifacts at fewer ACS lines (red arrows), due to the error propagation from the inaccurate sensitivity maps. By comparison, IMJENSE can produce reliable reconstruction image quality even with 4 ACS lines, as supported by the quantitative evaluation metrics in Fig. 7(b).

**Table 4.** Comparison of PSNR and SSIM obtained by different methods on the macaque brain dataset. The best results are in bold. Data are presented as Mean±Standard deviation.

|  | PSNR | SSIM |
|---|---|---|
| GRAPPA | 22.95±3.54 | 0.750±0.076 |
| JSENSE | 25.44±1.86 | 0.755±0.046 |
| NLINV | 27.76±1.81 | 0.827±0.041 |
| L1-ESPIRiT | 25.85±1.81 | 0.843±0.032 |
| LORAKS | 29.47±2.35 | 0.877±0.025 |
| SHLR-SV | 30.27±2.09 | 0.890±0.025 |
| IMJENSE | **31.71±0.90** | **0.899±0.021** |

### E. Comparisons on the Human Brain Dataset

Fig. 8 compares the reconstruction results of different methods on the human brain dataset with only 8 ACS lines. Visually, SHLR-SV, H-DSLR, and IMJENSE significantly improve the quality of reconstructed images at both R=5 and R=6, with IMJENSE exhibiting less noise in the results. In contrast, other compared methods fail to eliminate noise and artifacts due to highly limited ACS data. Quantitatively, IMJENSE achieves the highest PSNR of 33.87 dB at R=5 and 32.27 dB at R=6, as well as the highest SSIM of 0.940 at R=5 and 0.918 at R=6. Fig. 9 compares the k-space data recovered by different methods. Overall, the proposed method shows lower estimation errors in both low-frequency and high-frequency regions as illustrated by the error maps relative to the ground truth k-space. Fig. 10 plots the PSNR and SSIM variations of different methods as a function of the number of ACS lines at R=5 and R=6, respectively. H-DSLR and IMJENSE are more robust to ACS sizes compared to conventional methods. Particularly, IMJENSE achieves higher SSIM values across all ACS sizes.

Table 5 reports the quantitative evaluation metrics for the human brain dataset, which consists of 320 slices with 8 ACS lines. IMJENSE achieves the highest PSNR values of 35.00 dB at R=5 and 32.51 dB at R=6, as well as the highest SSIM values of 0.959 at R=5 and 0.942 at R=6.

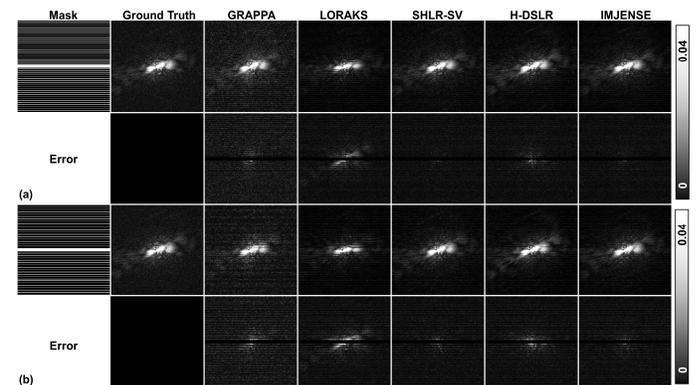

**Fig. 9.** The restored k-space and the corresponding errors relative to the fully sampled k-space on the human brain dataset with 8 ACS lines at (a) R=5 and (b) R=6. The magnitude k-space and errors from the first channel are displayed with a range of [0, 0.04] for better visualization.

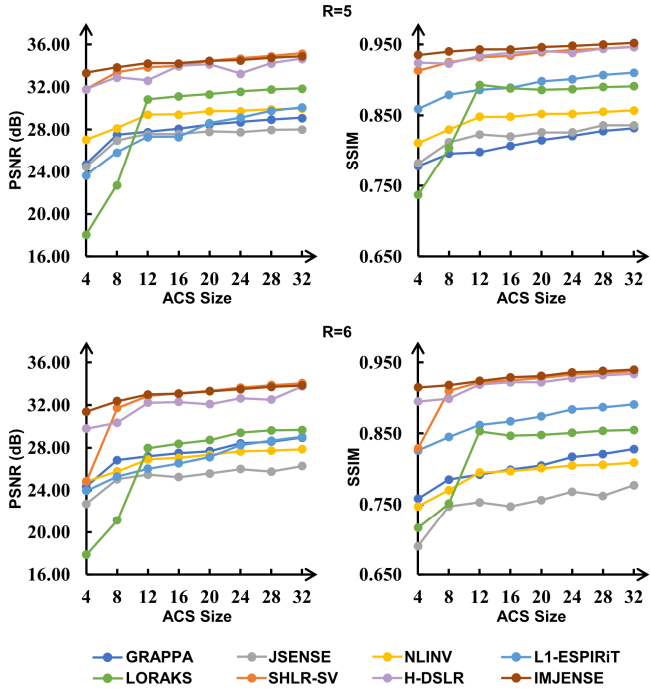

**Fig. 10.** Performance variations of different methods on the human brain dataset as a function of ACS sizes at R=5 and R=6, respectively.

**Table 5.** Comparison of PSNR and SSIM obtained by different methods on 320 slices from the human brain testing dataset with 8 ACS lines. The best results are in bold. Data are presented as Mean±Standard deviation.

| | R=5 | | R=6 | |
| --- | --- | --- | --- | --- |
| | PSNR | SSIM | PSNR | SSIM |
| GRAPPA | 31.36±1.90 | 0.903±0.039 | 28.87±1.90 | 0.882±0.042 |
| JSENSE | 27.17±2.05 | 0.866±0.028 | 25.68±1.35 | 0.825±0.032 |
| NLINV | 29.95±1.97 | 0.896±0.028 | 27.36±1.55 | 0.856±0.032 |
| L1-ESPIRiT | 26.36±1.72 | 0.906±0.021 | 24.91±1.72 | 0.881±0.025 |
| LORAKS | 26.71±1.22 | 0.898±0.019 | 25.29±1.23 | 0.869±0.023 |
| SHLR-SV | 33.91±1.32 | 0.941±0.017 | 31.45±1.45 | 0.921±0.021 |
| H-DSLR | 32.57±2.10 | 0.944±0.021 | 30.66±2.07 | 0.926±0.026 |
| IMJENSE | **35.00±1.27** | **0.959±0.011** | **32.51±1.54** | **0.942±0.014** |

### F. Comparisons on the Lesion Dataset

Fig. 11 displays the reconstruction results of the lesion data at R=4 using 8 ACS lines. The abnormal hyperintensities in the white matter are more noticeable in the reconstruction results of SHLR-SV, H-DSLR, and IMJENSE, as pointed out by the red arrows in the zoomed-in images. In contrast, GRAPPA and NLINV suffer from severe noise, making it difficult to distinguish the lesion region from the surrounding tissues. Moreover, the results of the proposed IMJENSE method show the lowest errors relative to the ground truth, with the highest PSNR and SSIM (32.65 dB and 0.934).

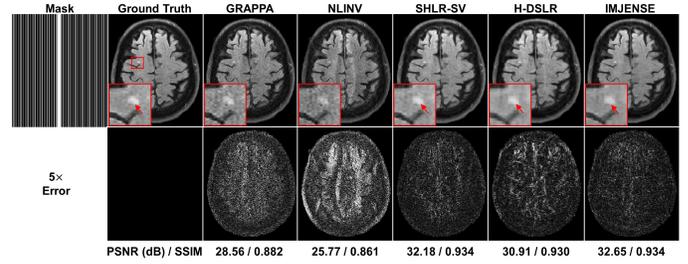

**Fig. 11.** Comparisons of different methods on the lesion dataset at R=4 with 8 ACS lines. The red box indicates the zoomed-in lesion region. Red arrows point to the lesion region that is more noticeable in the reconstruction results of SHLR-SV, H-DSLR, and IMJENSE. Quantitative evaluation metrics are reported below each image. IMJENSE achieves the highest PSNR and SSIM.

### G. Benefits over Calibration-based Methods

Fig. 12 illustrates the advantages of IMJENSE over the calibration-based MoDL method. With only 12 ACS lines, the image reconstructed by MoDL suffers from severe artifacts, as indicated by the red arrows, due to inaccuracies in the pre-estimated sensitivity maps. In comparison, IMJENSE produces excellent reconstruction results using both 24 and 12 ACS lines benefiting from its joint estimation scheme.

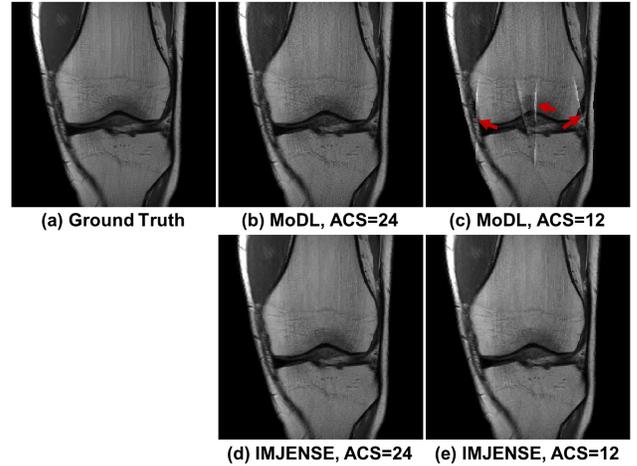

**Fig. 12.** Comparisons between (b-c) calibration-based supervised MoDL and (d-e) the proposed IMJENSE method at R=4, using 24 and 12 ACS lines, respectively. (a) The ground truth image. Red arrows indicate the severe artifacts in the MoDL results when only 12 ACS lines are used, due to inaccurate sensitivity maps estimated from a smaller calibration region.

### H. Results of the Ablation Experiments

Table 6 reports the quantitative evaluation metrics of different variants on the knee dataset. The proposed full model achieves the highest PSNR and SSIM in all cases, demonstrating the effectiveness of the SIREN network, TV loss function, and k-space consistency step in the current configuration.

### I. Time Consumption

Table 7 lists the mean computational speeds of each method on different datasets. For the knee, human brain, and lesion datasets, IMJENSE takes a comparable and even shorter amount of time than JSENSE and NLINV. Whereas for the macaque brain dataset, IMJENSE exhibits a slower computational speed due to the larger matrix size.

**Table 6.** Quantitative evaluation metrics of different variants in the ablation experiments on the 15-channel knee dataset. Data are presented as PSNR (dB) / SSIM.

| ACS size | Acceleration factor | IMJENSE-Sine | IMJENSE-Sine+PE | IMJENSE-TV | IMJENSE-KC | IMJENSE |
|---|---|---|---|---|---|---|
| ACS=24 | R=4 | 34.80/0.927 | 38.79/0.945 | 38.89/0.940 | 39.53/0.929 | 40.86/0.960 |
| | R=5 | 34.37/0.914 | 36.69/0.927 | 36.05/0.909 | 38.25/0.923 | 39.02/0.949 |
| ACS=32 | R=4 | 34.41/0.931 | 39.19/0.948 | 39.33/0.944 | 39.07/0.927 | 41.04/0.962 |
| | R=5 | 33.88/0.920 | 36.90/0.937 | 36.93/0.919 | 37.88/0.920 | 39.05/0.952 |

**Table 7.** Runtime comparison of different methods on the four datasets, in the unit of seconds, with the matrix size specified for each dataset. The device used for each method is also indicated.

| | Knee dataset (640×368) | Macaque brain dataset (710×840) | Human brain dataset (240×236) | Lesion dataset (320×320) |
|---|---|---|---|---|
| GRAPPA (CPU) | 8.0 | 12.3 | 2.9 | 2.1 |
| JSENSE (CPU) | 351.3 | 153.1 | 188.1 | 178.0 |
| NLINV (CPU) | 322.9 | 65.9 | 80.3 | 107.5 |
| L1-ESPIRiT (CPU) | 14.6 | 34.0 | 12.0 | 25.6 |
| LORAKS (CPU) | 46.7 | 30.2 | 34.3 | 15.1 |
| SHLR-SV (CPU) | 478.0 | 1293.2 | 470.5 | 486.2 |
| H-DSLR (GPU) | 0.4 | -- | 0.3 | 0.3 |
| IMJENSE (GPU) | 243.0 | 455.3 | 78.0 | 128.8 |
| IMJENSE +Hash (GPU) | 2.5 | 5.6 | 1.5 | 1.7 |

## VI. DISCUSSION

In this study, we proposed a novel scan-specific deep learning method, IMJENSE, for robust parallel MRI reconstruction by simultaneously learning the continuous representation of the MRI image and coil sensitivities. We validated IMJENSE on different datasets with various numbers of coil channels, acceleration factors, and ACS sizes. The results show that the proposed method can effectively suppress aliasing artifacts and noise, even at 5× and 6× accelerations with only 4 or 8 ACS lines, corresponding to 22.0% and 19.5% undersampling rates, respectively. These results demonstrate the promise of IMJENSE for further accelerating the MRI acquisition.

### A. Rationality of SIREN for an Image Model

The SIREN network is essentially an MLP. In theory, MLPs are universal function approximators capable of fitting arbitrary complex continuous functions [66]. The hyperparameter $w_0$ adjusts SIREN's fitting capacity for high-frequency information (see Section VI.B). Consequently, by tuning $w_0$, it is possible for SIREN to learn a continuous function representing the image, as the image information primarily resides in relatively lower frequency bands compared to noise and artifacts. To demonstrate that the trained SIREN network has indeed learned the continuity of MRI images, we queried the network in Experiment (1) with denser coordinate grids, resulting in 2×, 3×, and 4× upsampled images. Notably, in these denser grids, a significant portion of the coordinates were not encountered during the training stage. As shown in Fig. 13, as the upsampling scale increases, the image appears progressively smoother and more continuous compared to the original-sized image. These results suggest that after being trained on discretely sampled data, the network has encoded a continuous image function, rather than simply learning a pixel-by-pixel mapping from given coordinates to corresponding intensities. Consequently, SIREN implicitly imposes a continuity prior on the represented MRI image, allowing the proposed framework to obtain better solutions from the inverse problem of MRI reconstruction.

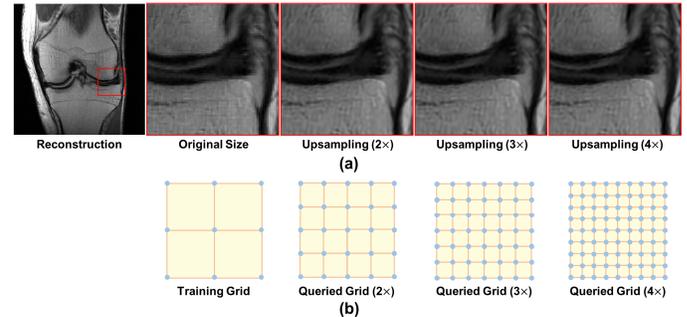

**Fig. 13.** (a) One representative knee slice of the reconstructed image. The region outlined by the red box is displayed with the original size, as well as 2×, 3×, and 4× upsampling obtained by querying the trained MLP using denser coordinate grids. (b) Illustration of the training grid and queried denser grids used in each case.

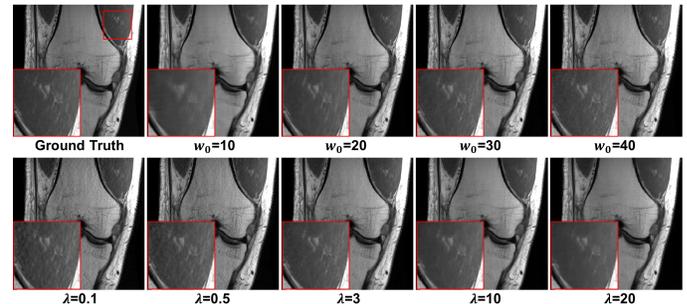

**Fig. 14.** Effects of $w_0$ and $\lambda$ on the reconstruction results.

### B.Effects of Hyperparameters in IMJENSE

In the proposed IMJENSE framework, $w_0$ and $\lambda$ are responsible for regulating the high-frequency information of the reconstructed image implicitly and explicitly. As shown in Fig. 14, an increased value of $w_0$ enhances the MLP's capacity to learn high-frequency information, which in turn produces more detailed textures in the reconstructed image. Conversely, a higher $\lambda$ results in smoother results. Therefore, it is necessary to fine-tune these two hyperparameters to achieve satisfactory reconstruction results. In this study, the selection of $w_0$ and $\lambda$ is dataset-specific, which means that for each dataset, only a small portion of the data is used to determine the parameters by the Bayesian optimization method. These parameters are then applied to the rest of the dataset. Quantitative statistical results on three types of datasets (Tables 3-5) indicate that this tuning strategy is efficient and ensures superior performance of the proposed method. The tuning of the two hyperparameters, $w_0$ and $\lambda$, is similar to selecting the regularization parameters in the optimization problem, such as in the compressed sensing reconstruction for MRI. These hyperparameters are always dependent on the dataset being used. Regularization parameters are estimated either through empirical means or by employing methods such as the L-curve approach. Other methods for more efficient selection of these hyperparameters based on the used data are of great interest for future research.

### C.Analysis of IMJENSE

The superiority of IMJENSE over the compared methods can be attributed to 1) the synergy between the internal continuous representation of an image offered by INR and the external regularization term (total variation) in the loss function; 2) the strategy for simultaneous estimation of coil sensitivity maps and the MRI image. Our experimental results indicate that IMJENSE outperforms the conventional joint estimation methods, e.g., JSENSE and NLINV, highlighting the value of implicit representation in solving inverse problems for parallel MRI reconstruction. The ablation results in Table 6 show that a large portion of the performance improvement can be attributed to the continuous representation, but the best results are obtained when both the continuous representation and the total variation are incorporated. Furthermore, the joint estimation scheme enables the proposed method to correct potential inaccuracies in sensitivity maps, thus leading to improved image reconstruction results, especially when the ACS size is extremely small (4 or 8), as illustrated in Figs. 7, 8, and 12.

Accurate coil sensitivity estimation is particularly crucial for explicit image domain reconstruction methods. Therefore, significant efforts have been dedicated to jointly optimizing coil sensitivities and the MRI image. Based on the traditional optimization algorithm, JSENSE initially estimates coil sensitivities using the ACS data and then alternates between updating coil sensitivities and the MRI image [42]. Although some deep learning methods have also adopted the joint estimation strategy, they often optimize sensitivities only from the low-frequency k-space data, typically the fully sampled ACS region [36, 40, 41, 44]. Recently, researchers have started using all measured k-space data for more faithful sensitivity estimation [37]. Similarly, the proposed IMJENSE method also estimates coil sensitivities based on the entire measured k-space. Benefiting from the gradient back-propagation algorithm, the sensitivity maps can be optimized simultaneously with the MRI image by starting from a random initialization, making IMJENSE a highly flexible framework that can be applied to the reconstruction with both calibration-based and calibrationless sampling patterns.

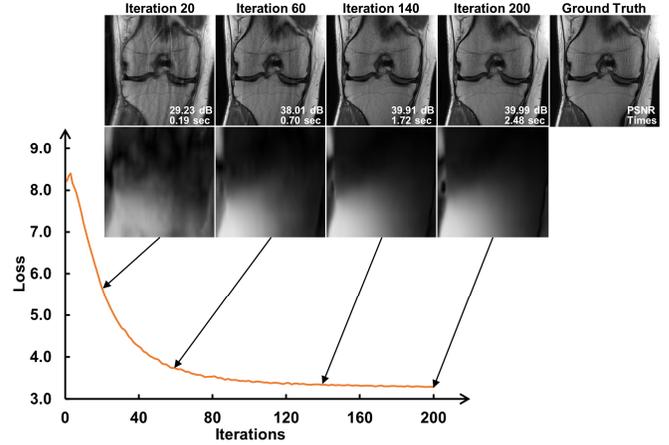

**Fig. 15.** The convergence of the accelerated framework using hash encoding on the FastMRI knee dataset at R=5 with 24 ACS lines. The bottom row shows the training loss at each iteration. The top row shows the reconstruction results and runtime at different iterations. The model takes only 2.48 seconds for the reconstruction of a knee slice with a matrix size of 640×368, producing results comparable to the current framework.

### D.Limitations and Future Work

The current IMJENSE requires a longer computation time than GRAPPA, L1-ESPIRiT, LORAKS, and H-DSLR methods (Table 7). To further speed up the reconstruction of our proposed framework, we exploited the utilization of hash encoding [55]. The continuous functions of coil sensitivities and the MRI image were both parameterized using hash encoding and an MLP with one hidden layer, while keeping the other parts of the framework unchanged. The parameters of hash encoding used to represent the MRI image remained their default values, while those used to represent the sensitivity maps were configured as follows: the number of multi-resolution levels $L$=8, the size of each hash table $T$=2, the dimension size of each feature vector $\log_2 F$=18, the coarsest resolution $N_{min}$=1, and the increasing factor of the resolution $b$=1.5. A detailed description of these parameters can be found in [55]. Particularly, we reduced the coarsest resolution $N_{min}$ from the default value of 16 to 1 due to the low-dimension nature of coil sensitivities, which results in smooth sensitivity maps and thus ensures high-quality MRI image reconstruction. Fig. 15 displays the reconstruction results and runtime on the FastMRI knee dataset at R=5 with 24 ACS lines. The accelerated method achieves better reconstruction results than the current IMJENSE framework within 2.48 seconds (39.99 dB *vs.* 39.02 dB in Fig. 3(c)). The reconstruction speed is approximately two orders of magnitude faster than the current framework. Additionally, the runtime of IMJENSE implemented through hash encoding (IMJENSE+Hash) on different datasets is summarized in Table 7. The test code can be found at: https://github.com/AMRI-Lab/IMJENSE. These preliminary experimental results demonstrate the potential of the proposed implicit representation framework to be further accelerated. Our future work will conduct a comprehensive evaluation of hash encoding on larger datasets. Additionally, there are other acceleration strategies that researchers can explore, such as exploiting the meta-learning algorithm to learn the initial weights of MLP for a class of signals being represented [67] or using transfer learning to apply the learned weights to the reconstruction of other data. These techniques facilitate faster convergence when handling a batch of new data. It is worth noting that all our current tests were conducted on an NVIDIA GeForce RTX 3090 GPU. When it comes to clinical applications, higher-performance GPUs, parallel computing using multiple GPUs, and tailored hardware accelerators [68] can be employed for further acceleration. These aspects, involving engineering considerations, are beyond the scope of this paper.

## VII. Conclusion

We have demonstrated a novel deep-learning insight that fundamentally differs from previous deep learning-based parallel MRI reconstruction methods. The proposed IMJENSE is a training database-free method and implicitly learns the continuous function representation of the MRI image and coil sensitivities from the partially acquired k-space itself. We tested IMJENSE on different datasets under different conditions. Results show that IMJENSE can improve parallel MRI reconstruction with highly reduced k-space measurements in the 2D Cartesian acquisition. The superior performance and the scan-specific characteristic make the proposed method potential for further speeding up the MRI data acquisition.